\documentclass{iau-JDSS}
\usepackage{graphicx}

\title[Magnetic reconnection in jet/accretion disk systems] %% give here short title %%
{On the role of magnetic reconnection in jet/accretion disk systems}

\author[de Gouveia dal Pino et al.]   %% give here short author list %%
{Elisabete M. de Gouveia Dal Pino$^1$,%
\break Pamela Piovezan$^{1, 2}$,  Luis Kadowaki$^1$, \break Grzegorz Kowal$^1$ \and Alex Lazarian$^3$}

\affiliation{$^1$ IAG, Universidade de S\~ao Paulo, Rua do Mat\~ao 1226,  S\~ao Paulo 05508-900, Brazil \break dalpino@astro.iag.usp.br \\[\affilskip]
$2$ Karl-Schwarzschild-Str. 1, Postfach 1317, D-85741 Garching, GermanyMPA, Garching, Germany\\[\affilskip]
$^3$Astronomy Department, University of Wisconsin, Madison, WI, USA }

\pubyear{2010}
\volume{15}
\pagerange{}
\date{May}
\jname{IAU Highlights of Astronomy, Volume 15}
\editors{Ian F. Corbett et al., ed.}
\begin{document}

\maketitle

\begin{abstract}
The most accepted model for jet production is based on the magneto-centrifugal acceleration out off an accretion disk that surrounds the central source (\cite[Blandford \& Payne, 1982]{bp82}). This scenario, however, does not explain, e.g.,
 the quasi-periodic ejection phenomena often observed in different  astrophysical jet classes.
 \cite{bl2005} (hereafter GDPL) have proposed that the large scale superluminal ejections
observed in microquasars during radio flare events could be produced by violent magnetic reconnection (MR) episodes.
Here, we extend this model  to other accretion disk systems, namely: active galactic nuclei (AGNs) and young stellar objects (YSOs), and also discuss its hole on jet heating and particle acceleration.

%% add here a maximum of 10 keywords, to be taken form the file <Keywords.txt>
\end{abstract}

\firstsection % if your document starts with a section,
              % remove some space above using this command.
\

\textbf{MR IN MICROQUASARS AND AGNS:}  A violent MR process between the magnetic field lines of the accretion disk  and
those that are anchored into the black hole may occur when a large scale magnetic field is established by turbulent dynamo in the inner disk region with a ratio between the gas+radiation and the magnetic pressures $\beta \leq 1$. During this process, substantial angular momentum is removed from the disk by the wind generated by the  magnetic flux, which increases the disk mass accretion rate to a value near the Eddington limit. After the reconnection, the partial destruction of the magnetic flux in the inner disk will make it to return to a less magnetized condition with most of the energy being dissipated locally within the disk instead of in the outflow.
The magnetic power released by MR (see Figure 1) is able to heat the coronal/disk gas and
accelerate the plasma to relativistic velocities through a diffusive first-order
Fermi-like process within the reconnection site that will produce intermittent
relativistic ejections or plasmons (GDPL). The resulting power-law electron distribution is
compatible with the synchrotron radio spectrum observed during the outbursts of these
sources. We are presently testing this acceleration mechanism  with  fully 3D numerical simulations.
The diagram of the magnetic energy rate released by violent reconnection as a
function of the black hole (BH) mass spanning $10^9$ orders of magnitude (Figure 1)  shows that
the magnetic reconnection power is more than sufficient to explain the observed radio
luminosities of the outbursts, from microquasars to low luminous AGNs (LINERs and Seyfert galaxies).
 This result is consistent with  recently found empirical relation that correlates the observed radio emission from   microquasars and radio quiet AGNs to that of magnetically active stars (Laor \& Behar 2008; Soker \& Vrtilek 2009),   suggesting  that it is  mainly due to  magnetic activity in the coronae and therefore, is nearly independent of the intrinsic physics of the central source and the accretion disk.
The correlation found in Figure 1 does not hold for radio-loud AGNs, possibly because their surroundings are much denser and then "mask" the emission due to   coronal magnetic activity. In this case, particle re-acceleration behind shocks further out in the jet launching region will be probably the main responsible for the radio emission. The violent MR could also be  responsible for the transition from the so called 'hard' steep power-law state (SPLS) to the 'soft' SPLS in microquasars (\cite[Remillard \& McClintock, 2006]{rm2006}).

\begin{figure}[t]
\begin{center}
\includegraphics[scale=.5]{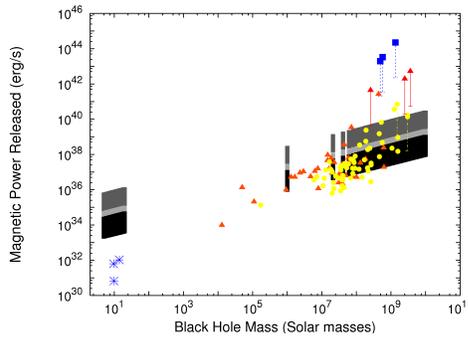}
\caption{Magnetic power due to violent reconnection versus the BH mass for both microquasars and AGNs. The stars represent the observed radio luminosities for  three microquasars.  The circles, triangles and squares are observed radio luminosities of jets at parsec scales from LINERS,  Seyfert galaxies, and luminous AGNs, respectively.  The thick bars correspond to the calculated
magnetic reconnection power and encompass a fiducial parameter space
(see de Gouveia Dal Pino, Piovezan \& Kadowaki, 2009, for details).}
\label{fig:all}       % Give a unique label
 \end{center}
\end{figure}

\textbf{MR IN YSOS:} The observed flares in x-rays are often attributed to magnetic activity at the stellar corona. However, some COUP (Chandra Orion Ultra-deep Project) sources have revealed strong flares that were related to peculiar gigantic magnetic loops linking the magnetosphere of the central star with the inner region of the accretion disk. It has been argued that this x-ray emission could be due to magnetic reconnection in these gigantic loops (\cite[Favata et. al, 2005]{f2005}).
We have extended the MR scenario described above to these sources and
found that a similar magnetic configuration
can be reached that could possibly produce the observed x-ray flares in most of the sources and provide
the heating at the jet launching base if violent magnetic reconnection events occur
with episodic, very short duration accretion rates  $\sim 100-1000 $ times
larger than the typical mean accretion rates expected for more evolved (T Tauri) YSOs.

\begin{acknowledgments}
The authors acknowledge partial support from FAPESP and CNPq.
\end{acknowledgments}

\end{document}